\documentclass[useAMS,usenatbib]{mn2e}

\usepackage{graphicx}
\usepackage{bm}
\usepackage{color}

\title[Betelgeuse Hybrid Wind]{The Magnetised Bellows of Betelgeuse}
\author[Anand Thirumalai and Jeremy S. Heyl]{Anand Thirumalai$^{1}$\thanks{E-
mail:anand@phas.ubc.ca (AT); heyl@phas.ubc.ca (JSH)} and Jeremy S. Heyl$^{1}$
\footnotemark[1]\\
$^{1}$University of British Columbia, 6224 Agricultural Road, Vancovuer, British Columbia, 
V6T 1Z1, Canada}

\begin{document}

\date{\today}

\pagerange{\pageref{firstpage}--\pageref{lastpage}} \pubyear{2011}

\maketitle

\label{firstpage}

\def\aj{AJ}                   % Astronomical Journal
\def\apj{ApJ}                 % Astrophysical Journal
\def\apjl{ApJ}                % Astrophysical Journal, Letters
\def\aap{Astronomy and  Astrophysics}                % Astronomy and Astrophysics
\def\araa{ARA\&A}             % Annual Review Astronomy and Astrophysics
\def\apjs{ApJ}                % Astrophysical Journal Supplement Series
\def\apss{ApSS}		%Astrophysics and Space Science
\def\mnras{MNRAS}             % Monthly Notices of the RAS
\def\nat{Nature}              % Nature
\def\physrep{Phys.~Rep.}      % Physics Reports
\def\pra{Phys. Rev. A} 			%Physical Review A
\def\pre{Phys. Rev. E} 			%Physical Review E
\def\prb{Phys. Rev. B} 			%Physical Review B
\def\prd{Phys. Rev. D} 			%Physical Review D
\def\pasp{Pubs. Astron. Soc. Pac.}
\def\solphys{Sol. Phys.}

\begin{abstract}
We present calculations for a magnetised hybrid wind model for Betelegeuse ($\alpha -
$Orionis). The model is a direct application of our previously derived theory, combining a 
canonical Weber-Davis (WD) stellar wind with dust grains in the envelope of an AGB star  
\citep[see][]{Thirumalai2010}. The resulting hybrid picture provides a mechanism for solving 
the problem of lifting stellar material up from the photosphere \citep[e.g.][]
{Harper2009,Guandalini2006,Jura1984} and into the circumstellar envelope. It also predicts 
wind velocities in agreement with current estimates. Our approach reveals that magnetic 
fields in supergiant stars like Betelgeuse \citep[see][]{Auriere2010}, may play a vital role in 
determining the nature of the stellar outflow and consequently, opens a new avenue of 
investigation in the field of hybrid stellar winds. 
\end{abstract}


\section{Introduction}\label{sec:intro}
Betelgeuse ($\alpha -$Orionis), one of the closest cool-evolved supergiant stars to Earth, has 
been the focus of research for many decades now. However, despite the considerable 
amount of attention it has received, our understanding of the wind of the star and the nature of 
its 
mass efflux remain as yet, mysterious in more ways than one.  Even though modern understanding of 
the stellar wind of this M-type supergiant star, despite formidable challenges on both the 
observational and theoretical fronts, has progressed well beyond the rudimentary stages 
\citep[see][]
{Holzer1985,David1990,Danchi1994,Smith2009,Harper2006,Harper2008,Plez2002, Bernat1979,Noriega1997,Ueta2003}, still however, significant gaps in our understanding remain 
quite 
entrenched. For example, observations in neither the ultra-violet (UV) nor the infra-red (IR), 
reveal blue-shifted emission lines; a requisite signature of gas outflow \citep[see][]
{Harper2009}, should the wind get sufficiently accelerated close to the star.  It is also still not known 
how the stellar wind is supported. We do however understand that dust forms at small and large distances 
from 
the photosphere, at radii $r \stackrel{_>}{_\sim} 1.33R_0$
\citep[e.g.][]{Perrin2007}
and $r \stackrel{_>}{_\sim} 25-30R_0$ \citep[e.g.][]{Bester1991,Bester1996,Danchi1994,Harper2009,Skinner1997}
respectively where $R_0$ is the optical photospheric radius. While there is now evidence of dust forming species such as alumina being present in small quantities, in the lower reaches of Betelgeuse's atmosphere ($ \sim 1.3 - 1.5 R_0$) \cite[see][]{Perrin2007,Verhoelst2006}, there is no evidence that this results in a dust-driven wind at these distances \cite[e.g.][]{Harper2009}. It has been conjectured that alumina may form close to the star and then, once transported to large distances ($r \stackrel{_>}{_\sim} 30R_0$), may provide nucleation sites for silicate dust to form \cite[see][]{Verhoelst2006}. However, this scenario still requires transport of stellar material to these larger distances in the atmosphere and presently, it is 
not known how this is achieved \citep
[see][]
{Harper2009,Guandalini2006,Jura1984}. It is believed that a combination of MHD effects, 
pulsation and convection may be responsible, but there are presently no models that 
demonstrate this unequivocally \citep[see][]{Harper2009}. On the other hand, efforts at 
modelling the stellar wind of Betelgeuse (and red supergiants and AGB stars) remain almost as disparate 
factions, modelling either solely magnetohydrodynamic (MHD) or acoustic waves \citep[e.g.][]
{Hartmann1980,Hartmann1984,Falceta2002,Falceta2006} or relying upon a combination of 
dust- 
\citep[see][]{Elitzur2001,Liberatore2001,Woitke2005} and pulsation-driven mechanisms \citep[see][]
{Elitzur2003,Bowen1988,Bowen1991,Fleischer1992,Wood1979,Bedijn1988,Gail1988}. Each of these models achieves a modicum of success, while ignoring stellar rotation altogether. 
Though it is widely acknowledged, that both magneto-centrifugal effects and the presence of 
dust grains in the gas, may be part of a greater coupled picture \citep[see][]
{Thirumalai2010,Guandalini2006,Falceta2002,Falceta2006,Schroder2007}, there have 
however been no efforts, thus far, at combining the two for a red supergiant like Betelgeuse. 

In addition to these concerns, it is also clearly seen that the atmosphere around Betelgeuse is 
not spherically symmetric. Indeed, there are some inhomogeneities that are seen, attributable to clumps in the outflow \citep[see][]{Plez2002}. These concerns are compounded by the fact that observations reveal 
that the temperature structure has a complicated form. Differences are seen at the same radius in different parts of
the atmosphere \citep[see][]{Lim1998}, revealing a slight departure from a spherically symmetric picture. With regard to seeing a clear signature of a stellar 
wind in the form of blue shifted emission lines, the observations have had severe difficulties in 
being able to resolve the spectra at close distances from the photosphere \citep[see][]
{Harper2009}. Indeed, observations have revealed that in parts of Betelgeuse's atmosphere 
there is redshifted emission indicating that there may be infall of matter back onto the star \citep[see][]{Lobel2001}.

The rotation rate of Betelgeuse has also been reported in the literature \citep[e.g.][]
{Uitenbroek1998}. More recently, Zeeman observations also reveal unambiguously, that 
there exists a magnetic field on the surface of Betelgeuse \citep[see][]{Auriere2010, 
Petit2011}. These observations naturally raise several questions regarding the role that 
magneto-rotational effects might play in the star's wind. Thus overall, we see that there are 
several unanswered questions regarding the nature of the outflow from Betelgeuse. Here, we 
present the very first model integrating MHD and rotational effects with a dust-driven wind 
scenario for Betelgeuse. The aim of the current study is to shed a new light on these issues 
and attempt to answer, at least some of these concerns. In this study, we find that the presence of a small magnetic field, on the order of what was recently 
discovered; of about $1$~G  \citep[see][]{Auriere2010}, is sufficient to drive material from 
close to the stellar surface up and out of its gravity well, by means of a magneto-rotational wind. Dust condensation later occurs at large distances from the star ($\sim 30R_0$) which finally results in  
a hybrid-MHD-dust-driven outflow. This model provides a possible alternative resolution of this issue \citep[see][]
{Harper2009,Guandalini2006,Jura1984}. It is to be mentioned at the very outset, that this model is merely suggested as an \emph{additional} mechanism that can play a role in transport of stellar material and is not intended to supplant the altogether feasible models involving MHD or acoustic waves as likely candidates. 

The theoretical 
model that we developed for intermediate mass asymptotic-giant-branch (AGB) stars \citep
[see][]{Thirumalai2010} has been extended to tackle the case of a supergiant like Betelgeuse. 
The reader is referred to our earlier work \citep[see][]{Thirumalai2010} for details of the model. 
Here we shall present only the salient features of the theory. We also find that the wind 
velocities that are obtained from the model are in good agreement with current estimates 
\citep[e.g.][]{Harper2009,Falceta2002,Plez2002}. 

\section{The Hybrid-Wind Model}\label{sec:model}

For the purposes of modelling, there are some marked differences that Betelgeuse exhibits from a run-of-the-mill AGB star. For example, Betelgeuse is much more massive ($\sim 15 \mathrm{M}_{\odot}$, \citealt{Smith2009}) than an 
intermediate mass AGB star. It also has a far more extended and cooler atmosphere. More importantly, the dust condensation radius is much farther out, in terms of stellar radii, from the photosphere in comparison to a canonical AGB star. In the former, the primary process governing mass-loss is the radiation pressure on dust grains that form at close distances from the photosphere, typically no greater than $10 R_0$, coupled with strong stellar pulsation. The presence of a magnetic field then has the effect of playing a secondary role governing the gas dynamics. In an AGB star, the hybrid-MHD-dust-driven mechanism is such, that without the onset of dust formation, it is not possible to achieve an outflow and thereafter, the predominant energy exchange in the wind is between the magneto-rotational and gravitational components \cite[see Fig.~5 of][]{Thirumalai2010}. In such a scenario, it was seen quite crucially, that dust formation must occur prior to the hybrid model's sonic point and concomitantly, it was seen that the dust parameter, $\Gamma_d$ was required to be less than unity for achieving a hybrid wind. Such is not the case for Betelgeuse where observations find only a small amount of dust inside $ r \stackrel{_<}{_\sim} 25-30~R_0 $.

In our earlier study, we had additionally investigated the possibility of locating the dust condensation radius outside the fast Alfv\'{e}n point. It was shown that such a hybrid wind is entirely theoretically possible, given typical parameters of an AGB star \cite[see Fig.~9 of][]{Thirumalai2010}. It was concluded therein that while such a scenario is unlikely for a typical AGB star it may well apply to an altogether different type of stellar wind. It is this second type of hybrid-MHD-dust-driven model that is adapted herein to formulate a stellar efflux scenario for Betelgeuse, with dust formation occurring at a large distance $\sim 30~R_0$, from the star.

Our theoretical model can be summed up as follows (the interested reader is referred 
to \citet{Thirumalai2010} for the steps involved in the derivation). We imagine that we are 
looking down upon the two-dimensional equatorial plane of Betelgeuse. Therein, we assume 
that the magnetic field and the gas (and dust) velocity are functions of purely the radial 
distance from the centre of the star. The poloidal (co-latitudinal) components of these 
vectors vanish; this is the fundamental assumption behind the canonical Weber-Davis \citep
[see][hereafter WD]{WD67} model for our sun. The gas forms the first fluid and carries the 
magnetic field. However, unlike the sun, in the atmosphere of Betelgeuse, embedded within 
the gas is a second fluid; the dust. The two fluids co-exist and are coupled to each other 
through drag. In an evolved supergiant like Betelgeuse, the circumstellar atmosphere is cool 
enough ($\stackrel{_<}{_\sim}1000$K) that dust grains can condense out of the surrounding gas \citep[e.g.][]{Bester1991,Bester1996,Danchi1994,Harper2009,Skinner1997}. In fact, direct imaging has enabled estimates for the inner dust shell temperature to be $\sim 700~$K at around $\sim 30R_0$ \citep[e.g.][]{Bloemhof1984}. Stellar radiation from the interior impinging upon the dust grains, can impart 
enough momentum to power these exiguous solar sails and propel them outward through the 
surrounding gaseous matter. However, whilst moving through the gas, the dust \emph{drags} 
the gas along with it, resulting in a prodigious and combined mass efflux of both dust and gas, 
from the star; the so-called dust-driven wind.  It is to be kept in mind, that the dust-to-gas mass-ratio is small; for Betelgeuse, it is expected to be on the order of $\approx 6\times 
10^{-4}  - 5 \times 10^{-3}$ or so \citep[see][]{Harper2001}. Additionally, the dust 
grains in our model are assumed to be spherical in shape, thus presenting a circular cross-section for radiation pressure to act upon. The surface temperature of Betelgeuse ($T_0$) 
plays a key role in determining the hybrid wind parameters, such as the bulk gas velocity at the base of the wind and the location of the critical points (see below). We also 
assume that the gas has the thermodynamic equation of state of a polytrope, with a polytropic 
index $\gamma > 1$, where a value of unity represents the isothermal limit. We model the 
fluid flow as an inviscid one and the electrodynamic properties of the fluid are taken to obey 
ideal MHD, i.e., there is no Lorentz  force acting on the fluid and the electric and magnetic forces 
balance each other completely. Additionally, the model requires as input, a number of 
observed quantities, such as the mass of the star, its rotation rate, the surface magnetic field 
strength and the mass-loss rate. These and other parameters for Betelgeuse are listed in Table~1. With these ingredients, upon examining the Euler equations for fluid flow for 
both the dust and the gas that are coupled with each other and invoking mass and energy flux 
continuity and ensuring that the divergence of the magnetic field explicitly vanishes within the 
governing equations, we can arrive at a steady-state description of the hybrid dual-fluid wind 
in the equatorial plane of Betelgeuse \citep[see][]{Thirumalai2010}. The radial equation for 
the gas velocity profile is then given by,
\begin{eqnarray}
\frac{dw}{dx}=\frac{w}{x}\frac{N(w,x)}{D(w,x)}~,
\label{eq:1}
\end{eqnarray}
where, $w=u/u_A$ is the gas speed normalised using the Alfv\'{e}n
speed and $x=r/r_A$, is the radial distance expressed in units of the
Alfv\'{e}n radius. Hereafter, the subscript `$A$' refers to values of the different variables at the 
Alfv\'{e}n radius. The quantities $N(w,x)$ and $D(w,x)$ are the
numerator and denominator respectively and are given by,
\begin{eqnarray}
N(w,x)= \left(2 \gamma S_T (wx^2)^{1-\gamma} - \frac{S_G}{x}(1-\Gamma_d \cdot \Theta(x-
x_d))\right) \nonumber\\
\times (wx^2-1)^3 +~ S_{\Omega}x^2(w-1)\left(1-3wx^2 + (wx^2+1)w \right)
\label{eq:2}
\end{eqnarray}
and
\begin{eqnarray}
D(w,x)= \left(w^2-\gamma S_T (wx^2)^{1-\gamma}\right)(wx^2-1)^3-
S_{\Omega}x^2 \times \nonumber\\
(wx^2)^2\left(\frac{1}{x^2}-1\right)^2.
\label{eq:3}
\end{eqnarray}
In the above equations, the parameters $S_T=\frac{2kT_A}{m_p u_A^2}$,
$S_G=\frac{GM_*}{r_A u_A^2}$ and $S_\Omega=\frac{\Omega^2
  r_A^2}{u_A^2}$ along with $\gamma$ uniquely determine the locations
of the critical points, and hence the morphology of the family of
solutions of Eq.~(\ref{eq:1}). Here $T_A$ is the gas temperature at
the Alfv\'{e}n radius, $k$ is the Boltzmann constant and $m_p$ is the
mass of a proton. The critical points are, as usual, defined as the
locations at which both the numerator and the denominator vanish,
thereby keeping the right-hand side of Eq.~(\ref{eq:1}) finite, these
are the sonic point, the radial Alfv\'{e}n point and the fast point
\citep[e.g.][]{WD67}. The presence of the Heaviside function in
Eq.~(\ref{eq:2}) represents the formation of dust at the location
$x=x_d$, the dust condensation radius in units of the Alfv\'{e}n
radius. The critical wind solution of Eq.~(\ref{eq:1}) will yield the
gas velocity profile, thereby enabling the determination of all other
dependent variables, such as the dust velocity profile (to be
discussed below), the Mach number as a function of distance from the
star, the azimuthal velocity of the gas, the azimuthal component of
the magnetic field, the temperature profile and the density structure
of the gas in the envelope of the Betelgeuse. The critical solution of
Eq.~(\ref{eq:1}) is defined as one that starts off at the base of the
wind sub-sonic, passes through the three critical points in a
continuous manner and emerges super-Alfv\'{e}nic at large distances
from the star. The dust velocity profile is then given by \citep
[see][]{Thirumalai2010},
\begin{equation}
v(r)=u(r)+\left(\frac{\sqrt{a_{th}^4+4\left(\frac{\Gamma_d GM_{*}}{\pi a^2 n_d r^2}\right)^2}-a_
{th}^2}{2}\right)^{1/2},
\label{eq:4}
\end{equation}
where $a_{th}$ is the thermal speed given by $a_{th}=\sqrt{2kT/\mu m_u}$ and $\mu m_u$ is 
the mean molecular mass of the gas and $n_d$ is the dust grain number density, which is 
assumed to be given by, $n_d m_d / \rho \approx \langle\delta\rangle$ where $\langle\delta
\rangle$ is the average dust-to-gas ratio in the wind
\citep[see][]{Thirumalai2010}. The dust in the current theory is treated in a simplistic and idealised manner, without rigorously considering the effects of dust radiative properties or including the effects of scattering and absorption on the radiation pressure mean efficiency. While such an analysis would no doubt portray a more complete picture, the current rudimentary treatment nevertheless captures the salient features of the coupled outflow from the star. As our purpose here is to illustrate the feasibility of a hybrid-MHD-dust-driven wind model for Betelgeuse, the current simplistic treatment of dust was considered sufficient.

Eq.~(\ref{eq:1}) is solved numerically, and the reader is referred to our earlier work 
\citep[see][]{Thirumalai2010} for complete details on the numerical methodology. The pertinent points of the method are conveyed below, in brief.
\begin{table}
 \centering
\label{tab:Table1}
%\begin{minipage}{140mm}
  \caption{Various parameters for modelling Betelgeuse.}
  \begin{tabular}{@{}lcl@{}}
  \hline
   Parameter     & Symbol        &  Value / Comment \\
 \hline
Mass & $M_*$ & $\sim 15\mathrm{M}_{\odot}$ \\
Radius & $R_0$ & $\sim 650\mathrm{R}_{\odot}$ \\
Mass loss rate & $\dot{M}$ & $\sim 3 \times 10^{-6} \mathrm{M}_{\odot}$/yr \\
Surface magnetic & & \\ 
field strength & $B_0$ & $\sim 1$ G \\
Bulk surface gas & & \\ 
velocity (radial) & $u_0$ & $\sim 10^{-12} v_\mathrm{esc,0}$ \\
& & (vanishingly small) \\
Surface temperature (effective) & $T_0$ & $\sim 3650$K \\
Stellar rotation rate & $\Omega$ & $\sim 1.2 \times 10^{-8}$ rad/s \\
Surface escape velocity & $v_\mathrm{esc,0}$ & $9.39 \times 10^6$ cm/s \\
Polytropic exponent & $\gamma$ & $> 1$ \\
Alfv\'en Radius & $r_A$ & $\sim 25R_0$ \\
Alfv\'en speed & $u_A$ & $\sim 0.15 v_\mathrm{esc,0}$ \\
Dust Parameter & $\Gamma_d$ & varied \\
Dust grain radius & a & spherical grains \\
Dust grain mass & $m_d$ & $\sim 4/3 \pi a^3 \rho_d$ \\
\hline
\end{tabular}
%\end{minipage}
\end{table}

\section{Numerical Method}\label{sec:numer}
Eq.~(\ref{eq:1}) is solved using the package ODEPACK employing a finite difference method 
with chord iteration with the Jacobian supplied \citep[see][]{Hindmarsh1983,Hindmarsh1989}. 
Initial conditions were supplied
at the beginning of the integration. Typical error tolerances for
convergence testing that were employed were on the order of $10^{-12}$
for both the absolute and relative errors
\citep[see][]{Hindmarsh1983}. For a typical integration, step sizes of
$10^{-9}$ or $10^{-10}$, in units of the Alfv\'{e}n radius,
were employed depending upon the region of integration being near the critical points or 
sufficiently away from them. This 
resulted in typically $10^9-10^{10}$ function evaluations. In the
current study, in contrast to \cite{Thirumalai2010}, we located the
radial Alfv\'{e}n point at around $25R_0$ with an Alfv\'{e}nic
temperature of $T_A \approx 720$~K, and Alfv\'{e}nic velocity $u_A
\approx 0.15 v_\mathrm{esc,0} \approx 14~$km/s, because we wanted to
have a dust formation temperature of $\sim 700$~K with dust
condensation occurring at $\sim 30R_0$.  The polytropic exponent $\gamma$ was varied and the locations of the sonic point and the fast point were found according to the method described in \cite{Thirumalai2010}. Once a particular value for $\gamma$ is chosen, the bulk radial gas velocity at the photosphere is found using the relation,
\begin{equation}
u_0=u_A \left(\frac{r_A}{R_0}\right)^2 \left(\frac{T_A}{T_0}\right)^{1/(\gamma-1)}
\label{eq:5}
\end{equation}
Thus, with these parameters now identified, a solution of Eq.~(\ref{eq:1}) is obtained that is continuous through the critical points with a tolerance for testing continuity of about $10^{-8}$, about an order of magnitude greater than the integration step size. The polytropic exponent is varied until a suitable set of values for the parameters $\{r_s,u_s,r_f,u_f,u_0\}$, are obtained so that the solution would be first continuous through all the critical points and second, result in a temperature of about $700~$K in the vicinity of $30R_0$.
 
\section{Results and Discussion}\label{sec:results}
Observations of the circumstellar atmosphere of Betelgeuse
  indicate that dust primarily exists in the form of a shell at a
  distance of about $\sim 30 R_0$
  \citep[e.g.][]{Bloemhof1984}. However, recent
  observations as discussed above, seem to indicate that there may be
  small amounts of alumina (Al$_2$O$_3$) present rather close to the
  photosphere \citep[see][]{Perrin2007,Verhoelst2006}, within $1.5 R_0$. However, it is thought that this alumina may be either transient, perhaps even destroyed at around $1.5R_0$ or even further out in the chromosphere \citep[see][]{Verhoelst2006}, or is present in such small quantities as to be unable to support a dust-driven wind at these distances. There is a third possibility, however remote, that the small amount of alumina present does result in a mild dust-driven wind, but the alumina is transparent until it accumulates silicates on its surface \citep[see][]{Onaka1989}, which occurs in the dust shell at around $30 R_0$.

Regardless, observations do not indicate a significant presence of dust in the region $1.5 R_0 \stackrel{_<}{_\sim} r  \stackrel{_<}{_\sim} 30 R_0$. In this paper, we address each of these possibilities, within the framework of the hybrid-dust-driven wind model. Physically, there are four distinct scenarios that emerge and these are listed in Table 2.
\begin{table}
 \centering
\label{tab:Table2}
%\begin{minipage}{140mm}
  \caption{Different dust formation scenarios in Betelgeuse}
  \begin{tabular}{@{}ll@{}}
  \hline
%   Parameter     & Symbol        &  Value / Comment \\
% \hline
Scenario 1a  & Silicate dust forms at $r=30 R_0$ in reasonable\\ 
                     & quantities ($\Gamma_d=0.5$, $\langle \delta \rangle \sim 1/2000$)\\
                     & \\
Scenario 1b  & Silicate dust forms at $r=30 R_0$ in large \\ 
                     & quantities ($\Gamma_d=5$, $\langle \delta \rangle \sim 1/200$)\\
		     & \\
Scenario 2a & Alumina dust forms at $r  < 1.5 R_0$ in small \\
                    & quantities ($\Gamma_d=0.05$, $\langle \delta \rangle \sim 1/20000$) and \\
                    & provides nucleation sites for silicate dust \\
                    & around $r=30 R_0$ \\
		    & \\
Scenario 2b & Alumina dust forms at $r  < 1.5 R_0$ in large \\
                    & quantities ($\Gamma_d=0.5$, $\langle \delta \rangle \sim 1/2000$) and \\
                    & provides nucleation sites for silicate dust \\
                    & around $r=30 R_0$ \\
		    & \\		    
Scenario 3a & Alumina dust forms at $r  < 1.5 R_0$ in small\\
                    & quantities ($\Gamma_d=0.05$, $\langle \delta \rangle \sim 1/20000$) and \\
                    & is subsequently destroyed at $r \approx 7.53 R_0$ due \\
                    & to a chromospheric component in the atmosphere \\
                    & and silicate dust later forms at $r=30 R_0$ \\       
		    & \\
Scenario 3b & Alumina dust forms at $r  < 1.5 R_0$ in large\\
                    & quantities ($\Gamma_d=0.5$, $\langle \delta \rangle \sim 1/2000$) and \\
                    & is subsequently destroyed at $r \approx 7.53 R_0$ due \\
                    & to a chromospheric component in the atmosphere \\
                    & and silicate dust later forms at $r=30 R_0$ \\       
		    & \\
Scenario 4a & Alumina dust forms at $r  < 1.5 R_0$ in small\\
                    & quantities ($\Gamma_d=0.05$, $\langle \delta \rangle \sim 1/20000$) and \\
                    & is subsequently destroyed at $r \approx 1.5R_0$ and \\
                    & silicate dust forms at $r=30 R_0$ \\  
		    & \\
Scenario 4b & Alumina dust forms at $r  < 1.5 R_0$ in large\\
                    & quantities ($\Gamma_d=0.5$, $\langle \delta \rangle \sim 1/2000$) and \\
                    & is subsequently destroyed at $r \approx 1.5R_0$ and \\
                    & silicate dust forms at $r=30 R_0$ \\  
		    & \\
                                                               
\hline
\end{tabular}
%\end{minipage}
\end{table}
Each of these scenarios are explored in the following discussion.

\subsection{Scenario 1: Silicate dust forms at $30 R_0$}
\begin{figure}
\begin{center}
\includegraphics[width=3.5in, scale=0.9]{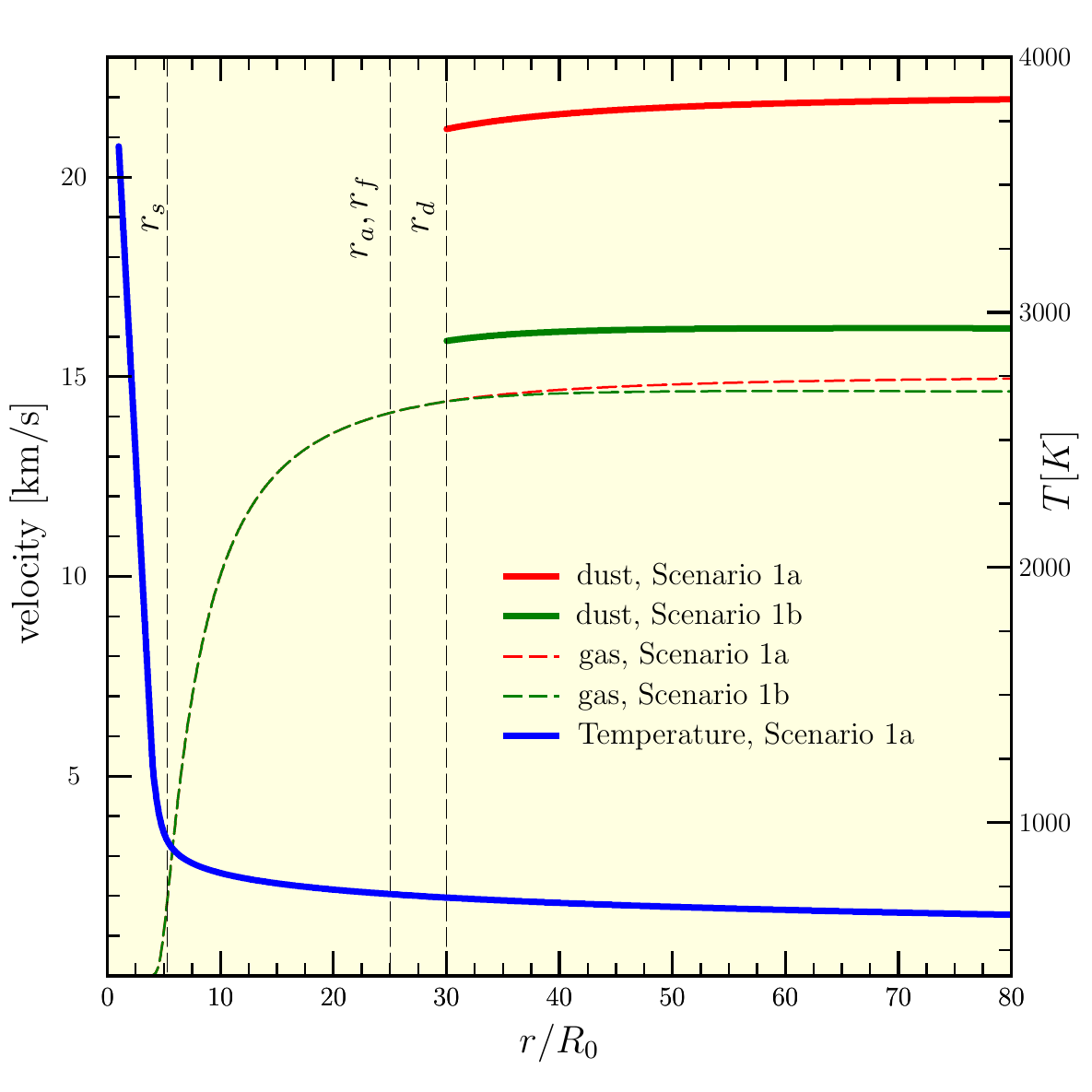}
\end{center}
\caption{A hybrid wind solutions are shown for Scenario 1 with parameters $u_A=0.15v_{esc,
0}$, 
$r_A=25R_0$, for different $\Gamma_d$ and remaining parameters as given in
Table 1.  The red and green solid lines trace the dust velocity profiles for Scenarios 1a and 1b respectively, corresponding gas velocity profiles are shown by the red and green long-dashed lines. The decreasing blue solid line
traces the temperature and should be interpreted using the right hand y-axis.} 
\label{fig:figure1}
\end{figure}

This is the simplest and perhaps the most likely scenario to fit the observations of Betelgeuse's atmosphere. Herein, there is no dust formation at close distances and silicate dust condenses at a distance of about $30 R_0$ where the temperature drops to about $700~$K. Stellar material is transported from the photosphere to this distance of $30 R_0$ by means of a Weber-Davis magneto-rotational wind. This is shown in Figure~\ref{fig:figure1}, with two cases corresponding to silicate dust formation in reasonable and large amounts at a distance of about $30R_0$.
The gas velocities of the critical hybrid-wind solutions of Eq.~(\ref{eq:1}) are shown in
Figure~\ref{fig:figure1} as the red and green long-dashed lines. These solutions start
at the base of the wind subsonic and  accelerate first through the
sonic point at about $5.27R_0$; after this the wind is supersonic. It
can be readily seen that the Mach numbers are small for the critical
solution close to the photosphere
($r\stackrel{_<}{_\sim}5R_0$). Beyond the sonic point, the wind mildly
accelerates through the radial and fast Alfv\'{e}n points, that are nearly coincident upon one another, and emerges super-Alfv\'{e}nic at large distances ($r > 25 R_0$). A little further out, dust condenses out from the gas
at a radial distance of $r_d = 30 R_0$, shown by the vertical dashed line marked ``$r_d$". At this distance from
the surface of Betelgeuse, the temperature has dropped below the dust
condensation temperature for silicates of about $\approx 700$~K; the temperature is given by
the blue solid line and should be interpreted using the right-hand axis. The temperature profile shown in Figure~\ref{fig:figure1}, corresponds to Scenario 1a, wherein silicate dust forms in reasonable amounts at around $30R_0$. Thereafter the wind is a coupled 
MHD-dust-driven wind and the gas outflow rapidly approaches the terminal velocity. Thus, the 
solution is purely WD from the phototsphere ($r=R_0$) to the dust condensation radius 
($r=r_d$) and thereafter it is hybrid. This is the primary salient feature of the hybrid wind 
model developed for Betelgeuse. 

Also shown in Figure~\ref{fig:figure1} are the corresponding dust velocities for these two models in this scenario, shown by the solid lines that lie above the long-dashed ones. It can be seen that the dust grains are moving radially faster than the
gas and dragging the gas along with them. The solid
red line represents the dust velocity profile for a model with parameters $\Gamma_d=0.5$
and $\langle \delta \rangle = 1/2000$; Scenario 1a. In order to investigate the effect of changing
the average dust-to-gas ratio and therefore the dust parameter $\Gamma_d$, we kept all other parameters of the model fixed,
in particular, the radiation pressure mean efficiency and the stellar luminosity,
were kept constant. Now, if the average dust-to-gas 
ratio is increased by an order of magnitude then accordingly, the dust parameter $\Gamma_d$ must also correspondingly increase by an order of magnitude. Thus, for the second model's results shown in Figure~\ref{fig:figure1}, we took $\Gamma_d=5$ and
$\langle \delta \rangle = 1/200$; Scenario 1b. For this latter model, the dust velocity profile is shown with the green solid line that lies below the red solid line. This range of the dust-to-gas ratio of $1/2000 \leq \langle \delta \rangle \leq1/200$, represents a reasonable bound for the amount of dust in the atmosphere of Betelgeuse \citep[e.g.][]{Harper2001}. 
It was seen in our earlier work \citep[c.f. Figure 6 of][]{Thirumalai2010} that changing the dust parameter shifted the location of the critical points. In general, increasing the value of the dust parameter $\Gamma_d$ results in moving the location of the sonic point and fast point towards the surface of the star; this is the case should dust formation occur inside the sonic point in the hybrid wind model. However, it was shown earlier \cite[see Figure 9 of][and discussion thereof]{Thirumalai2010} that formation of dust beyond the fast point does not influence the location of the critical points. Then, Eq.~(\ref{eq:1}) can simply be integrated with the presence of the Heaviside function from $r=r_d$ to $r=\infty$ for a given value of $\Gamma_d$. Thus in this case, the wind has already successfully passed through the critical points and emerged super-Alfv\'{e}nic prior to dust condensation.

Beyond about $r \approx r_A=25R_0$, the acceleration of the gas in the wind due to dust drag in the second model ($\Gamma_d=5$, Scenario 1b) starts to decline more steeply than in the case of a hybrid model with a smaller value of $\Gamma_d=0.5$ (Scenario 1a). Thus, the gas in the wind in the first model ($\Gamma_d=0.5$, Scenario 1a) at this distance, is still getting accelerated, therefore it's terminal velocity is slightly larger and the red long-dashed line lies above the green long-dashed line. Thus, when $\Gamma_d$ is smaller, acceleration due to radiation pressure continues to have an effect, out to larger distances from the star.

The effect on the dust grains is a little counter-intuitive and can be understood by examining Eq.~(\ref{eq:4}). We can re-write Eq.~(\ref{eq:4}) by replacing the dust grain number density with an expression employing the dust-to-gas ratio as,
\begin{equation}
v(r)=u(r)+\left(\frac{\sqrt{a_{th}^4+4\left(\frac{\Gamma_d GM_{*}m_d}{\pi a^2 \rho \langle \delta \rangle r^2}\right)^2}-a_
{th}^2}{2}\right)^{1/2}.
\label{eq:6}
\end{equation}
Upon examining the second term under the square-root; the radiation pressure term, we can see that the smaller the value of the average dust-to-gas ratio, the larger this term will be and therefore the larger the value of the dust grain velocity, $v(r)$. Thus, when the radiation pressure mean efficiency and the stellar luminosity are kept constant, then naturally, the dust grain velocity is larger for smaller dust-to-gas ratios; this is the effect seen in Figure~\ref{fig:figure1}. For the calculations carried out above, we assumed that the
dust grains were spherical and were assumed to be on
average \citep[see][]{Perrin2007} about 0.005$\mu$m in size with a density of
about 4 g/cm$^3$.

With regard to scattering of radiation by the dust grains, in the current study this was assumed to be absent, thus precluding the complications that arise upon including this effect. Briefly, the inclusion of isotropic scattering would have the effect, for the simple theory described here, of altering the radiation pressure mean efficiency as, $Q_{rp} \mapsto Q^A + Q^S$, where $Q^A$ and $Q^S$ represent the efficiencies of absorption and isotropic scattering, respectively. In a more rigorous model, these could be calculated for a particular type of dust grain and used in the equations, thus incorporating scattering of photons by dust grains. 

Moreover, in the framework of the current theory, the dust grains do not possess azimuthal velocity with respect to the gas. That being said, Poynting-Robertson drag due to scattering of radiation by dust grains would inevitably decelerate the grains in the azimuthal direction, thereby altering the momentum equations further. In reality however, it is to be acknowledged that scattering is probably anisotropic since the dust grains may well align themselves along field lines. Such a detailed analysis involving the complex phenomena touched upon above, while being extremely pertinent and closer to a realistic picture, was considered to be outside the scope of the current study, where the aim is to portray a simple picture.

It is implicitly assumed in our model that dust condensation occurs abruptly at a distance where the gas temperature ($T_{gas}$) falls below the dust condensation temperature $T_{dust}^{c}$. In addition, we assume that that the conditions are conducive for grain growth. In the current, rather idealised treatment of the dust, we are not concerned with the radiative properties of the dust. We are therefore content with the assumption that the temperature of the outflow is governed by the gas. This was considered to be reasonable considering that the dust-to-gas ratio employed was small $\langle\delta \rangle =1/2000$. It has also been shown that the temperature profile in the circumstellar environment of Betelgeuse is rather complicated \citep[e.g.][]{Rodgers1991}, with effects such as dust-grain drag contributing to heating. Such an analysis as well inclusion of rigorous dust radiative properties was considered to be outside the scope of the current paper. We simply assume for determining the dust-grain temperature, that the dust grains must be in radiative equilibrium with the stellar luminosity field and that it is optically thin. Therefore, we can assume a relationship for the temperature profile for the dust, outside the dust condensation radius as \citep[e.g.][]{Lamers,Onaka1989},
\begin{equation}
T_d(r) = C~T_0 \left(\frac{R_0}{2r}\right)^{2/5}, \forall ~ r \geq 30R_0,
\label{eq:7}
\end{equation}
where, the coefficient $C$, depends upon the radiation pressure mean efficiency. In the current study we are treating the dust in a rather simplistic manner, therefore instead of calculating the radiation pressure mean efficiency we can assume that the coefficient $C$ lies in a range between zero and unity, i.e., $0 < C \leq 1$. This then allows us to calculate a range of values for the dust temperature at the condensation radius, $T_d(30R_0)$. The mean of these calculated values, for the given range in $C$ and the adopted values of $R_0$ and $T_0$ given in Table 1, is found to be $ \left. \langle T_d \rangle \right|_{30R_0} \approx 390~$K. This rather rudimentary estimate is in reasonably good agreement with the value obtained by \citet[][]{Harper2001}, of $360$~K at around $33R_0$. The dust in the current model may therefore be able to mimic, in a qualitative sense, the radio and IR behaviour of the dust model of \citet[][]{Harper2001}. Given the idealistic nature of the current hybrid-MHD-dust-driven wind theory, the formulation of a detailed model, such as that described by \citet[][]{Harper2001}, was considered to be outside the scope of the current investigation. We shall end with a cautionary note, that a detailed analysis would nevertheless be required, before conclusions can be drawn about the dust related physics and we acknowledge this as a limitation of the current theory. 

In summary, the result in Figure~\ref{fig:figure1} demonstrates that it is possible to obtain a coupling between
magneto-centrifugal effects and the usual dust-driving
mechanism. There are however a few pertinent observations with respect
to the hybrid wind model in specific, that warrant mentioning.
First, it can be clearly seen, that magneto-centrifugal
effects, with a small magnetic field ($\sim 1$~G) and slow rotation,
$\Omega \approx 1.2 \times 10^{-8}$rad/s, \citep[see][]{Uitenbroek1998} can
quite clearly lift material from the photosphere of Betelgeuse up into
the circumstellar atmosphere. In the latter region, dust condenses
in the gas to result in a hybrid wind at large distances. This reveals that
magneto-centrifugal driving can be an additional mechanism for solving
the mystery of how to lift stellar material out into the circumstellar
envelope in not only Betelgeuse, but potentially in all cool, evolved
supergiant stars. Second, the temperature profile obtained in the
solution, is reasonably consistent with observations
\citep[see][]{Plez2002,Lim1998,Harper2009}. For example, the profile indicates that around
$2R_0$, the gas temperature drops to about $2840$~K, lying well within
the measured range \citep[see][]{Lim1998}. 

In addition, the temperature range in the region $R_0 \leq r \leq 4.1R_0$ can be inferred from Figure~\ref{fig:figure1} to be $1200 \stackrel{_<}{_\sim} T \leq 3650~$K and it can be seen that lower limit for the observed [Fe II] emission of $2110$~K \citep[see][]{Harper2009} lies in this region. It can also be seen that the gas velocities are small in this region, consistent with the lack of observations for Doppler blue-shifted wind signatures.

Third, the resulting terminal velocity is about $14-15$ km/s; this is consistent with present
estimates \citep[e.g.][]{Falceta2002,Jura1984} and in reasonably good agreement with the adopted value of about $\sim 10$ km/s for modelling by \citet[][]{Harper2009}. It is to be remembered that measured or inferred values for the outflow velocity would indicate a somewhat averaged value for gas velocity at a given distance from the star; however, within the framework of the current model, outflow velocity is calculated in the equatorial plane alone. Simulations of large-scale MHD convection in Betelgeuse have suggested that it is possible to have variation in the radial velocities on the order of $1-10~$km/s in both up- and down-flowing regions \cite[e.g.][]{Dorch2004}. In addition, it is also to be mentioned that the stellar efflux is thought to be variable in Betelgeuse, observations seem to suggest variability over a 40 year period or so \cite[see][]{Danchi1994}, however the current framework assumes a steady-state case, thus the interpretation of the calculated values of outflow obtained in this study, need to be tempered by these observations. In light of all this, we considered the agreement of velocities to be  good. 

In addition, once the radial velocity profile has been determined, it is possible to determine the azimuthal component of the magnetic field. With this it becomes possible to estimate, \emph{\'{a} posteriori}, the importance of the Lorentz force and any ion-neutral drag. The rationale being to check the validity of the ideal-MHD assumption. In the current framework, it is assumed implicitly that the ions and neutrals are well coupled in the equatorial plane and non-ideal MHD effects such as ambipolar diffusion are negligible and thus ignored. A way to assess this is to calculate the ratio of the gyrofrequency and the momentum exchange rate \citep[e.g.][]{Bai2011} as,
\begin{equation}
\beta_j (r) =\frac{Z_j e B (r) }{m_j c} \frac{(m+m_j)}{\rho(r) \langle \sigma v \rangle_j}.
\label{eq:8}
\end{equation}
Where $\beta_j$ is the ratio for the $j^{th}$ ionic species, $Z_j$ is the atomic number, $m_j$ is particle mass of the given ionic species, $B$ is the magnitude of the magnetic field, $e$ is the electronic charge, $c$ is the speed of light and $m$ is the averaged particle mass of the neutrals. Here $\rho$ is the gas density, determined once the solution to Eq.~(\ref{eq:1}) is calculated. Finally, $\langle \sigma v \rangle_j$ is the ion-neutral collision momentum transfer rate coefficient \citep[see Table 2.1 and Equation 2.34 of][]{Draine2011} and is given bya
\begin{equation}
\langle \sigma v \rangle_j = 2.0 \times 10^{-9} \left( \frac{m_H}{\mu}\right)^{1/2} ~ \textrm{cm}^3 ~ \textrm{s}^{-1},
\label{eq:9}
\end{equation}
where, $\mu = m_j \mu_n / (m_j + \mu_n) $ is the reduced mass in a typical ion-neutral collision, with $\mu_n$ the mean molecular mass of neutrals. Similarly, the electron-neutral collision momentum transfer rate can be determined as \citep[e.g.][]{Draine1983,Bai2011},
\begin{equation}
\left. \langle \sigma v \rangle_e \right|_r = 8.3 \times 10^{-9} \times \textrm{max} \left[ 1, \left(\frac{T (r)}{100~K}\right)^{1/2}\right]~ \textrm{cm}^3 ~ \textrm{s}^{-1}
\label{eq:10}
\end{equation}
Assuming that the largest single ionic species is Si, we can then calculate the ratio of the gyrofrequency to the momentum exchange rate, for both the ions ($\beta_{Si}$) and the electrons ($\beta_e$).

We find that $\beta_\mathrm{Si}^\mathrm{max} \approx 7.9
  \times 10^{-7}$, while $\beta_e^\mathrm{max} = 2.7 \times 10^{-4}$. In general, we find over the entire domain for $r$ in the equatorial plane, that $\beta_{Si}(r) \ll \beta_e(r) \ll 1$. As mentioned earlier, one the central assumptions in the model is that the Lorentz force vanishes in the fluid, i.e., we have force-free MHD, therefore it is to be expected that the ratios $\beta$ of the gyrofrequencies to the momentum transfer rate, would also accordingly be small. This then implies, that for the radial range considered in this study, i.e. $0 \leq r \leq 80R_0$, the Hall and ambipolar diffusion terms arising from electron-ion drift and ion-neutral drift respectively, are negligible. Additionally, the conductivity of the plasma can be estimated using the usual relation, $\sigma = 10^7 T_e^{3/2} \mathrm{K}^{-3/2} \Omega^{-1} \mathrm{cm}^{-1}$ \citep[e.g.][]{Vishniac2003}, where $T_e$ is the electron temperature, such that $T_e(r) > T_\mathrm{gas}(r)$. Figure~\ref{fig:figure1} shows that the minimum temperature for the gas far away from the photosphere is about $T_\mathrm{gas}^\mathrm{min} \approx 640~$K. We can therefore find an estimate for the lower limit of the conductivity using this temperature and estimate a value of $\sigma_\mathrm{min} \approx 1.6 \times 10^{11} ~ \Omega^{-1}~\textrm{cm}^{-1}$, yielding a magnetic Reynolds number \citep[e.g.][]{Davidson} of $\mathrm{Re}_m = U L \mu_0 \sigma_{min} \sim 10^{23}$, so Ohmic diffusion is unimportant  ($U \sim 10~$km/s and $L \approx R_0$ are typical velocity and length scales for Betelgeuse and $\mu_0$ is the permeability of the vacuum, $4\pi \times 10^{-7} \Omega \mathrm{s}/\mathrm{m}$). The assumption of effectively infinite plasma conductance, is then a reasonable assumption. Thus, we see that the central assumption of ideal-MHD in the equatorial plane of Betelgeuse is a reasonable first approximation to make, for the purpose of conveying a simple picture.

Finally, the Mach numbers in the inner
region of the circumstellar envelope are small and actually, do not
exceed unity before the sonic point, located at around $5.27R_0$. Therefore our model does not suffer from the same
drawbacks as many Alfv\'en wave models that have the artifact of having large Mach numbers
close to the surface of the star. Indeed, in our model the wind only becomes super-Alfv\'enic 
beyond the Alfv\'en point at around $25R_0$ and even then, only mildly so. 

\begin{figure}
\begin{center}
\includegraphics[width=3.5in, scale=0.9]{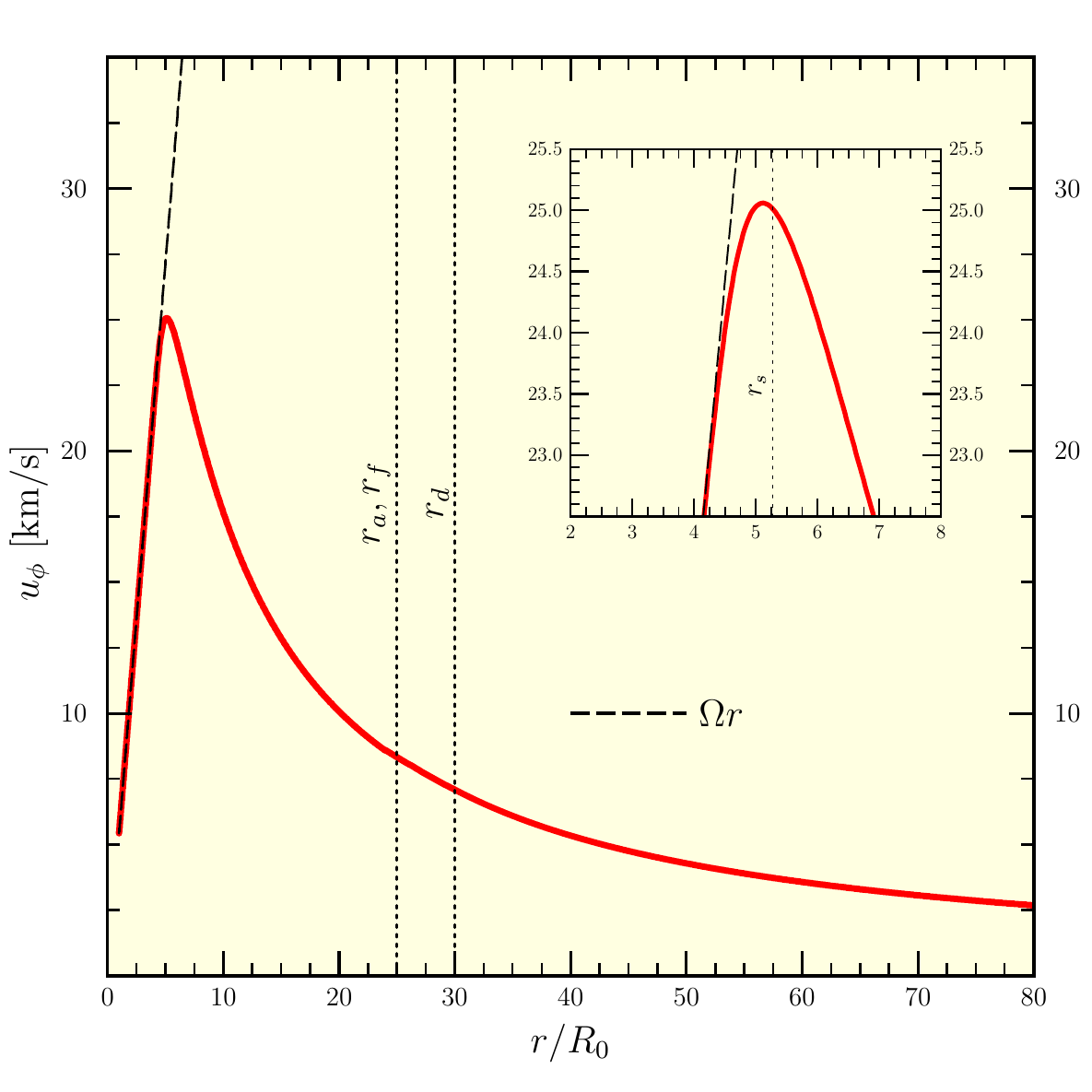}
\end{center}
\caption{The azimuthal velocity of the gas is shown as a function of the radial distance (red 
solid line) for Scenario 1a. The inset shows a magnified region around the peak of the profile around the sonic point.}
\label{fig:figure2}
\end{figure}

In Figure~\ref{fig:figure2} we have shown the azimuthal velocity, $u_{\phi}(r)$ of the gas (red 
solid line). The dust is assumed to co-rotate with the gas, thus the azimuthal velocity profile shown in the 
figure is also true for the dust (for $r \geq r_d$), since there is no drag between the dust and 
the gas in the azimuthal direction. The azimuthal velocity profile is obtained once the radial 
velocity profile is known \citep[see][]{Thirumalai2010}.  The velocity profile obtained is typical for a magneto-rotational wind. The dotted lines in 
the figure and in its inset represent the rotation of the star \citep[see][]{Uitenbroek1998}. It is 
readily seen that the azimuthal wind velocity closely traces the observed rotation of the star 
and only begins to depart markedly after $r \approx 4.5 R_0$. This decoupling from stellar rotation occurs in the inner wind region of the magneto-rotational wind, at a physical distance corresponding to $ \approx 2.03 \times 10^{14}$cm, for Betelgeuse. It is also interesting to note that at large distances the azimuthal velocity of the gas is strongly de-coupled from the stellar rotation rate. Therefore, the chromospheric component in the atmosphere at a few stellar radii may also appear to be de-coupled from stellar rotation as was observed by \cite[][]{Harper2006}. However, it is to be remembered that the hybrid-MHD-dust-driven theory presented in this study, concerns itself with the equatorial plane alone, while the observations of \citet[][]{Harper2006} are spatially resolved in the cross-dispersion direction; this may have an averaging effect along the dispersion direction. Thus, the theoretically obtained rotational decoupling that occurs in the equatorial plane, in the current study, may well be quite a bit different from the observations of \citet[][]{Harper2006}. Therefore any inferences drawn in this regard must be tempered by the observation that the current model is a rather simple picture, with concomitant limitations.

We now turn our attention to the question of spots on the surface
of Betelgeuse and the related question of temperature
inhomogeneities. Recent observations
\citep[see][]{Wilson1992,Wilson1997,Buscher1990,Haubois2009} suggest that
there may be a few (on the order of 2 or 3) large spots on the surface
of Betelgeuse, indicating that there are temperature inhomogeneities
on the surface of Betelgeuse. We therefore chose to model Betelguese
with two spots on its equator \citep[see][]{Soker1999}. 

In the current study however, these spots were taken to be colder than the effective temperature so that the temperature profile close to the photosphere $1.0 \leq r \leq 3.5 R_0$ may be investigated so as place constraints on the alumina formation region, if at all, close to the star. 
\begin{figure}
\begin{center}
\includegraphics[width=3.5in, scale=0.9]{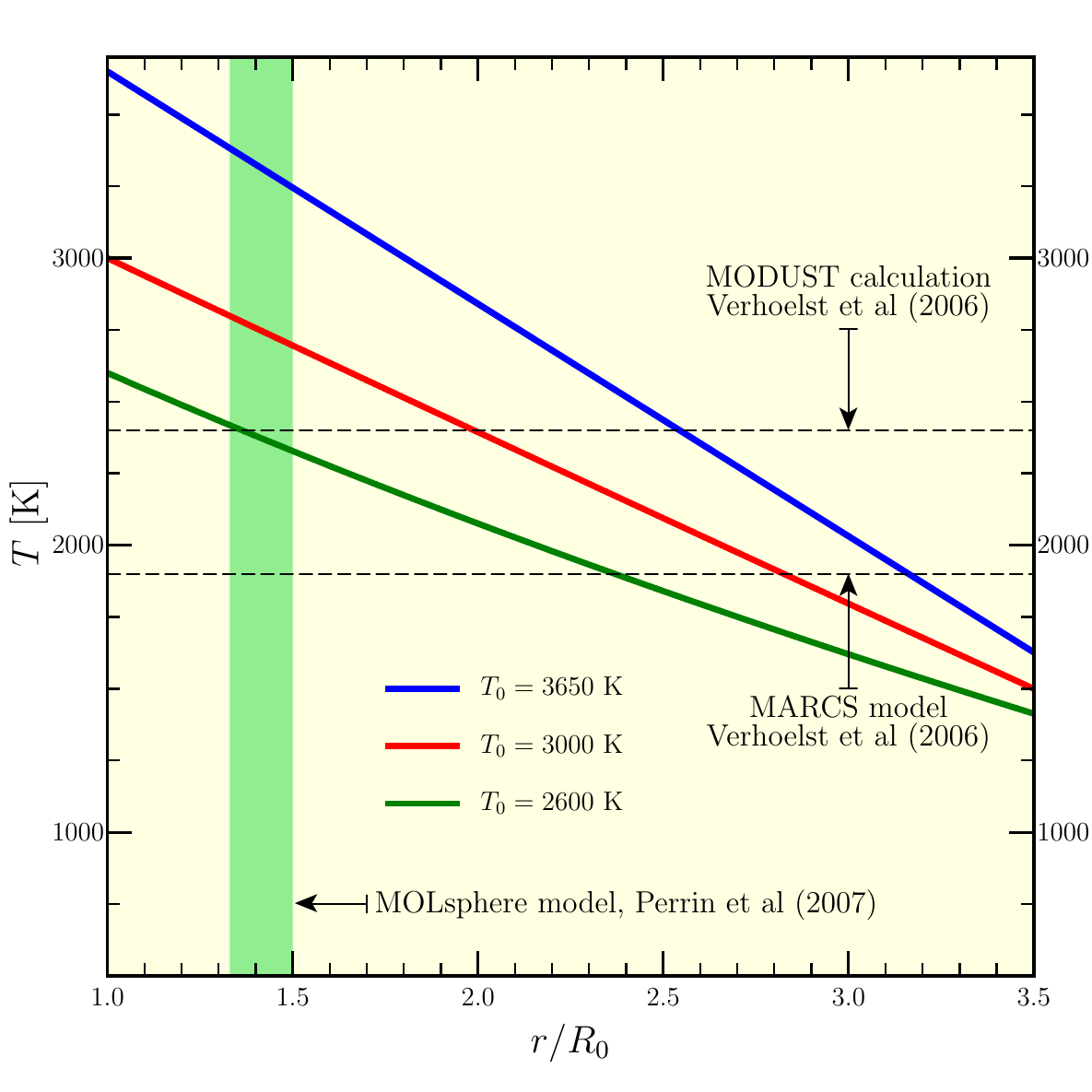}
\end{center}
\caption{Temperature profile in the close circumstellar environment with two photospheric spots of temperatures $2600~$K and $3000~$K. The two dashed horizontal lines show calculated alumina condenstation temperatures from two different models. The vertical green shaded region shows the MOLsphere region predicted by Perrin et al (2007). Only the model with photospheric spot temperature of $T_0=2600$~K has an overlap with the MOLsphere region for the alumina condensation temperature range.}
\label{fig:figure3}
\end{figure}
The spots were considered to have temperatures of $2600$~K and $3000$~K
respectively, well within fluctuations of about $1000$~K or so, about the effective temperature
\citep[see][]{Schwarzschild1975}. The magnetic field was left unchanged so
that the effect of changing a single parameter could be patently
established. The solitary requisite constraint that was placed however, was that regardless of the photospheric spot temperature, the temperature at a distance of about $30R_0$ should be in the vicinity of $700~$K, for silicate dust formation in the circumstellar shell at this distance. Thus, the radial Alfv\'{e}n point was left unchanged. Therefore the other free parameter that was varied in order to achieve an efflux was the bulk gas radial velocity at the surface, $u_0$. The gas velocity profiles were then solved for by integrating Eq.~(\ref{eq:1}), 
above the
spots located at the photosphere. Figure~\ref{fig:figure3} shows the temperature profiles obtained in the wind ahead of the spots in the close circumstellar environment of Betelgeuse. The temperature profile obtained using the effective temperature of $3650~$K is also shown for comparison as the top most blue solid line. It can be seen that the lower the photospheric temperature, the flatter the temperature profile is, in the inner wind region. Thus the red solid line ($T_0=3000$~K) lies below the blue solid line ($T_0=3650$~K) and above the green solid line ($T_0=2600$~K).	

Also shown in Figure~\ref{fig:figure3} are two temperature bounds calculated by \citet[][]{Verhoelst2006} for alumina dust condensation. The upper bound of $2400~$K is the value that they arrived at by demanding radiative equilibrium of dust grains using the radiative transfer code MODUST, for alumina condensation. The lower bound of $1900~$K on the other hand, is the temperature that their MARCS model predicts in the region about $0.5R_0$ above the photosphere. It is to be kept in mind however, that at such short distances, the MARCS model may have limitations due to acoustic wave heating and the breakdown of the assumption of local thermal equilibrium. With this caveat in place, it can be seen that this range of temperature only occurs in the region $2.55 R_0 \leq r \leq 3.16R_0$ for the model with $T_0=3650~$K, indicating the possible alumina dust condensation region. Similarly, the corresponding range for the model with the spot of temperature $T_0=3000K$ places this region to be $1.99 R_0 \leq r \leq 2.83R_0$. Both these regions lie well outside the observed thin shell for the MOLsphere of $1.33 R_0 \leq r \stackrel{_<}{_\sim} 1.5R_0$ \citep[e.g.][]{Perrin2007,Verhoelst2006}; shown as the green band in Figure~\ref{fig:figure3}.

It is however interesting to note, that the model for the photospheric spot temperature of $2600~$K, predicts a possible alumina condensation region of $1.37R_0 \leq r \leq 2.37R_0$. This region quite interestingly, overlaps slightly with the MOLsphere model of \citet[][]{Perrin2007}. However, it is to be remembered that current observations do not reveal any dust in the region $1.5R_0 \leq r \leq 20-30R_0$ \citep[e.g.][]{Verhoelst2006}. Thus, we see that within the framework of the hybrid-MHD-dust-driven model presented in this paper for Betelgeuse, it is possible to form alumina dust very close to the photosphere in the region of interest; $1.33 R_0 \leq r \stackrel{_<}{_\sim} 1.5R_0$, by lowering the photospheric temperature to about $2600~$K. For temperatures above this, our model predicts the possible alumina condensation region to lie further out from the photosphere and there is no overlap between our model and the MOLsphere model of \citet[][]{Perrin2007}. 

Additionally, it can be seen in Figure~\ref{fig:figure3}, that temperatures predicted by the models with photospheric spots are different at the same height above the photosphere, in the close circumstellar environment of Betelgeuse. Thus, the observed variability in
temperature \citep[see][]{Lim1998} at the same radial distance from the
photosphere can, at least in part, be attributed to the presence of
spots. The temperature profile in the wind above the spots is
different from that above normal regions. Thus, our model with
equatorial spots, can reproduce these differences in temperature,
at least qualitatively, indicating an asymmetric temperature distribution. 

Presently, we discuss the other possible scenarios for dust formation listed in Table 2.

The motivation for these scenarios is the observation that alumina dust may be present in small quantities close to the photosphere of Betelgeuse. Therefore, one can naturally ask the question whether alumina first forms in sufficient amounts to facilitate a slow dust-driven wind and second, what would happen if alumina dust is subsequently destroyed at some distance from the star? In this regard there are a few possibilities. It is to be kept in mind that there are no observations that support the presence of dust between $ \approx 1.5R_0$ and about $20-30 R_0$ \citep[e.g.][]{Verhoelst2006}. Therefore, the first possibility is we can assume that alumina forms in both small and sufficient quantities close to the star (\emph{Scenarios 2a and 2b} respectively), to facilitate a slow wind, but the alumina dust is transparent until it accumulates silicates on its surface beyond $30R_0$. In both these models, the assumption is that alumina dust provides nucleation sites for silicates to form upon.

The second possibility is that alumina dust forms close to the star (in small and large quantities; explored in \emph{Scenarios 3a and 3b} respectively). This results in a slow MHD-dust-driven hybrid wind. The dust is then destroyed at around $\approx 7.53R_0$ due to a chromospheric component in the atmosphere where the temperatures are high \cite[e.g.][]{Lobel2001}. Note that in \cite[][]{Lobel2001} they adopted a value for the radius of the photosphere as $700 R_{\odot}$, whereas we have adopted $650 R_{\odot}$, hence we obtain the outer limit for the chromosphere as $\approx 7.53 R_0$ rather than $7R_0$. Silicate dust later condenses at large distances $ \sim 30R_0$. It is implicitly assumed in this model that alumina dust is transparent so as not to reveal any dust signature between $1.5R_0$ and $30R_0$.

The third possibility is we can assume formation of alumina in the region $r \leq 1.5R_0$ (in small and large quantities; explored in \emph{Scenarios 4a and 4b} respectively). This dust then gets destroyed at a distance of $1.5R_0$ due to perhaps convective turbulence in regions closer to the photosphere and changes in pressure or perhaps due to temperature variability in the chromosphere. Whatever the reason may be, dust is not seen between $1.5R_0$ and about $20-30R_0$. Silicate dust later condenses at large distances. 

The purpose of these scenarios is not to elaborate on the details of dust spallation, but rather to ask the pertinent question that, within the framework of the hybrid-MHD-dust-driven wind theory, is it possible to achieve an efflux, should alumina first form and perhaps even be destroyed in the wind, at some distance? In addition, it is to be kept in mind that for the models presented in Scenarios $2-4$ the photospheric temperature was considered to be $T_0=2600~$K.

\subsection{Scenarios 2 and 3: Alumina forms and has an influence on the wind}
\begin{figure}
\begin{center}
\includegraphics[width=3.5in, scale=1]{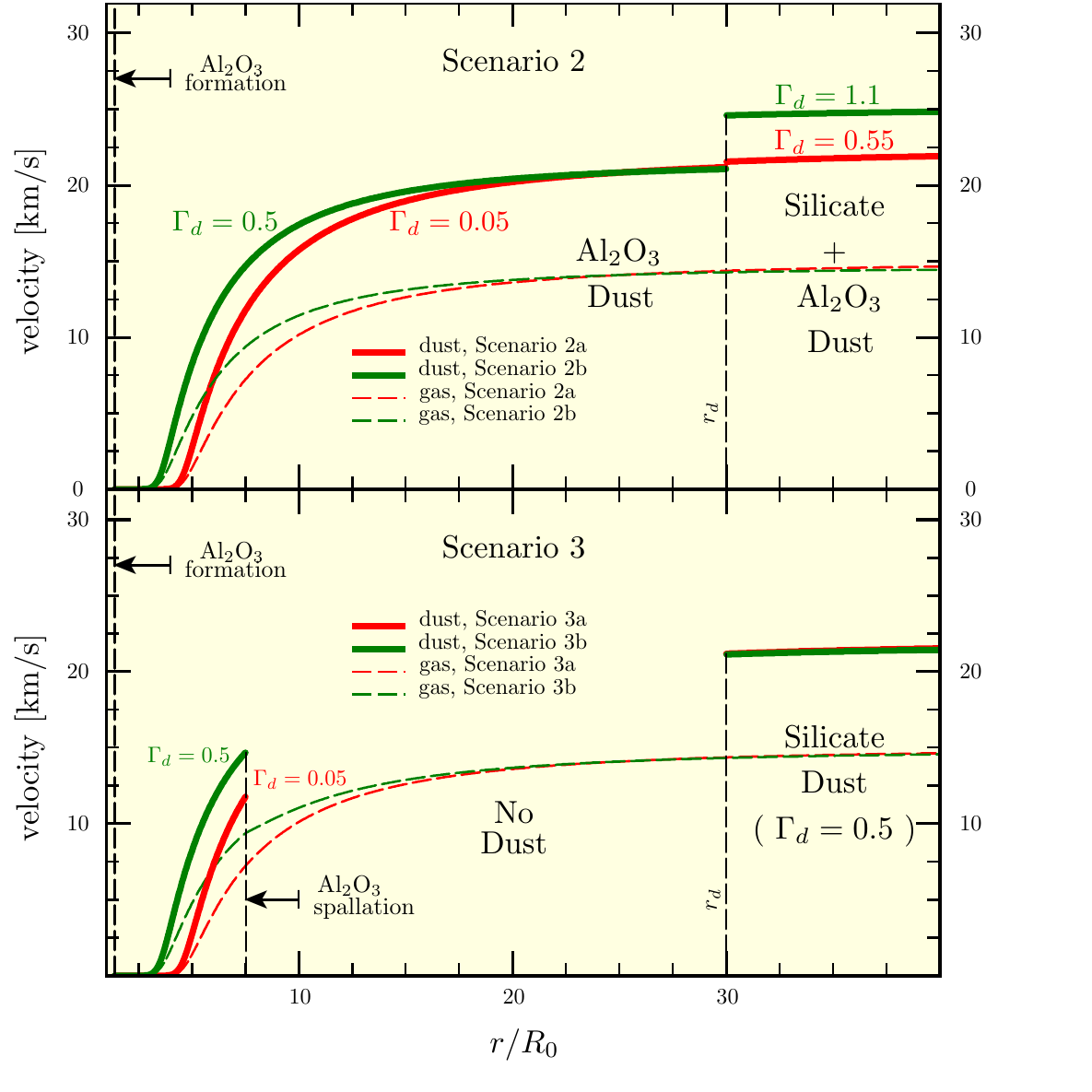}
\end{center}
\caption{Dust and gas velocity profiles for Scenarios 2 (upper panel) and 3 (lower panel). The dust velocities are shown using solid lines and the gas velocities with long-dashes lines. The dust and gas velocity profiles for the models with greater dust formation, i.e. larger $\Gamma_d$, lie above those with lesser amount of alumina dust formation in the inner wind region. Scenario 2 shows influence of the presence of both alumina and silicate dust species in the wind, whereas, Scenario 3 shows the effect of alumina spallation at the edge of the chromosphere at $7.53R_0$.} 
\label{fig:figure4}
\end{figure}

Figure~\ref{fig:figure4} shows both Scenarios 2 and 3 each containing two hybrid wind models. Scenario 2 is shown in the upper panel of Figure~\ref{fig:figure4} while Scenario 3 is shown in the lower panel. The red and green solid lines represent the dust velocity profiles in these scenarios for small and large amounts of alumina condensation, respectively. Similarly, the long-dashed green and red lines represent the corresponding gas velocity profiles. Quite importantly, in Scenarios 2a and 2b, the numerator in Eq.~(\ref{eq:2}) was modified to accommodate for the two different dust species; alumina and silicates as,
\begin{eqnarray}
N(w,x)= \left(2 \gamma S_T (wx^2)^{1-\gamma} - \frac{S_G}{x}\right.\nonumber\\
\left.\times(1-\Gamma_d^\mathrm{alumina} \cdot \Theta(x-
x_d^{(1)}) - \Gamma_d^\mathrm{silicate} \cdot \Theta(x-
x_d^{(2)}))\right) \nonumber\\
\times (wx^2-1)^3 +~ S_{\Omega}x^2(w-1)\left(1-3wx^2 + (wx^2+1)w \right),
\label{eq:11}
\end{eqnarray}
where, $x_d^{(1)}=1.38$ for alumina dust condensation and $x_d^{(2)}=30$ for silicate dust condensation. Thus, Eq.~(\ref{eq:1}) is integrated as a pure WD wind in the region $R_0 \leq r \leq 1.38R_0$, then with alumina dust in the region $1.38R_0 \leq r \leq 30R_0$ and finally with both alumina and silicate dust in the region $30R_0 \leq r \leq 40R_0$. In Scenario 2 (upper panel of Fig.~\ref{fig:figure4}) the portion of the hybrid solution in the region $1.38R_0 \leq r \leq 30R_0$ is a critical solution as it navigates through the critical points. The portions of the solutions outside this interval are not part of the critical solution, but rather lie elsewhere in the $u-r$ phase space of each of the models, respectively.

The alumina condensation radius of $1.38R_0$ lies in the range suggested by \citet[][]{Perrin2007}, for the thin molecular shell that contains alumina in the close circumstellar environment of Betelgeuse. This value also lies inside the green shaded region of Figure~\ref{fig:figure3}. The lower red solid lines in Figure~\ref{fig:figure4}, in both the upper and lower panels, represent the dust velocity profiles for models with formation of alumina dust in small quantities ($\langle \delta \rangle =1/20000$ ). This is an order of magnitude less than what is expected for silicate dust in the dust shell at around $30R_0$. Accordingly, we assumed an order of magnitude smaller value for the dust parameter, $\Gamma_d=0.05$; it is to be mentioned that even smaller values may be adopted. 

The upper green solid lines in both panels, prior to the silicate condensation region, represent the dust velocity profiles for models with formation of alumina dust in large quantities ( $\langle \delta \rangle =1/2000$). The dust parameter was then given by $\Gamma_d=0.5$. The purpose being to explore a somewhat extreme case in which alumina forms in large enough amounts to sustain a mild MHD-dust-driven hybrid wind, but is transparent to observations. While such a scenario may well be remote, we explore it for the sake of completeness.

It can clearly be seen in both Scenarios 2 and 3, as expected, that if the dust parameter is larger, then it results in a greater acceleration of the wind closer to the photosphere. Thus, the green lines lie above the red ones. For smaller values of the dust parameter, on the other hand, the effect of this acceleration due to dust, continues to have an effect out to larger distances. As a result, the red long-dashed lines rise above the green long-dashed lines at some distance. Ultimately the terminal velocity of the models with the smaller value of $\Gamma_d$ are slightly larger due to this effect.

In the upper panel in Figure~\ref{fig:figure4}, at the silicate dust condensation radius of $30R_0$, we notice a discontinuity, since the silicate dust condenses as well. The red solid line in this region corresponds to a model with parameters $\Gamma_d=\Gamma_d^\mathrm{alumina}+\Gamma_d^\mathrm{silicate} = 0.05 + 0.5 = 0.55$ with $\langle \delta_\mathrm{total} \rangle =1/2000$. Thus Eq.~(\ref{eq:1}) is integrated outwards from $r=r_d$ with these parameters. Similarly the green solid line in this region in the upper panel, represents a model with parameters corresponding to $\Gamma_d=\Gamma_d^\mathrm{alumina}+\Gamma_d^\mathrm{silicate} = 0.5 + 0.6 = 1.1$, again with $\langle \delta_\mathrm{total} \rangle =1/2000$. Notice that we have kept the dust-to-gas ratio fixed in both cases to the same value. The silicate dust condenses from the gas abruptly at $30R_0$. As a result, the discontinuity in the dust velocity across the silicate dust condensation radius is seen in both models. However, the jump in the velocity is much larger for the green solid line in comparison to the red solid line as this model has a higher value of $\Gamma_d$ for the same $\langle \delta_\mathrm{total} \rangle$. 

On the other hand, in the lower panel of Figure~\ref{fig:figure4} showing scenarios 3a and 3b, alumina dust is assumed to form close to the photosphere, but is then destroyed at a distance of about $7.53R_0$ due to a chromospheric component. There are some rather subtle differences from Scenarios 2a and 2b and these are detailed below.

In the lower panel of Figure~\ref{fig:figure4}, once again the lower solid red line shows the dust velocity profile for a model with parameters $\Gamma_d^\mathrm{alumina}=0.05$ and $\langle \delta_\mathrm{alumina} \rangle =1/20000$. The upper green solid line meanwhile, represents the dust velocity profile for a model with parameters $\Gamma_d^\mathrm{alumina}=0.5$ and $\langle \delta_\mathrm{alumina} \rangle =1/2000$. As described earlier, the solutions start off at the base of the wind at the photosphere, subsonic and alumina dust formation occurs at $r=1.38R_0$. Thus from $r=R_0$ to $r=1.38R_0$, Eq.~(\ref{eq:1}) is integrated as a pure WD wind without the dust parameter. It is to be mentioned that in this region the solutions are not part of the pure WD critical solution. Alumina dust condensation radius is shown at $1.38R_0$ for the two models in this scenario. As can be seen, the model with the larger dust parameter has a greater velocity, evidenced by the green lines lying above the red ones.

In the region $1.38R_0 \leq r \stackrel{_<}{_\sim} 7.53R_0$ the winds are hybrid MHD-dust-driven winds and Eq.~(\ref{eq:1}) is integrated with the presence of dust parameters for each of the two models. In this region the solutions for the gas velocities are parts of the critical solutions for each of the two hybrid-MHD-dust-driven winds. Alumina dust spallation is assumed to occur at a distance of about $\approx 7.53R_0$ and this produces the sharp discontinuity in both the dust and gas velocity profiles seen in the lower panel of Figure~\ref{fig:figure4}. In the region $7.53R_0 \stackrel{_<}{_\sim} r \leq 30R_0$, Eq.~(\ref{eq:1}) is therefore integrated without the dust parameter, i.e. as a pure WD wind, for the two models. However, the gas velocities in this latter region, are no longer part of the respective critical solutions. Hence, these solutions pass through only the radial Alfv\'en point, but not the fast point. Thus at large distances, these solutions are sub-Alfv\'enic. Therefore they constitute the so-called failed wind solutions as $r \rightarrow \infty$. As a result these models become theoretically non-viable as one of the primary requirements is that the wind should be super-Alfv\'enic at large distances. Thus with alumina dust spallation in between the sonic point and the radial Alfv\'en point, at around $7.53R_0$, we see that it is not possible to sustain a physically viable efflux.

For the sake of completeness, we have included silicate dust in Scenario 3. At a distance of $30R_0$ in the wind, silicate dust condensation occurs and beyond $30R_0$, the wind is once more a hybrid MHD-dust-driven wind. The dust velocities are once again calculated after the gas velocity profiles are determined, according to Eq.~(\ref{eq:4}). It can be seen that the gas outflow velocity profiles for the two models cross one another at around the radial Alfv\'{e}n point. As a result the dust and gas velocity profiles in the region $r \geq 30R_0$, for the model with greater amount of alumina (green dashed and solid lines, Scenario 3b) lie slightly below the one with the lesser amount of alumina (red dashed and solid lines, Scenario 3a). In both these models it is assumed that silicate dust forms with a dust-to-gas ratio of $ \langle \delta_\mathrm{silicate} \rangle =1/2000$, with $\Gamma_d=0.5$. 

In summary we see in the upper panel of Figure~\ref{fig:figure4}, that it is possible to obtain a hybrid-MHD-dust-driven wind if alumina forms close to the photosphere and results in a mild wind transporting stellar material to large distances. Silicate dust condenses at a distance of $30R_0$ and adds to the stellar efflux. Inspecting the upper panel of Figure~\ref{fig:figure4} reveals that the model with a larger amount of alumina formation close to the star (Scenario 2b), is not viable, since observations do not indicate such a large amount of alumina close to the photosphere. In the second model (Scenario 2a) with a smaller value of $\Gamma_d$, while the amount of dust may well be reasonable, current observations do not reveal any dust velocities. The only way we can reconcile this model with observations is by assuming that the alumina dust is transparent. 

In the lower panel of Figure~\ref{fig:figure4} on the other hand, once again it is seen that it is possible to have alumina formation at close distance, within $1.5R_0$ and subsequent spallation at around $\approx 7.53R_0$ and still achieve a solution with the formation of silicate dust. However, as stated earlier, the gas velocity solutions are not part of the critical solution. As a result, these solutions are sub-Alfv\'enic as $r \rightarrow \infty$. Therefore, they cannot be considered to be viable wind solutions. As a result, alumina spallation at a distance of $7.53R_0$ does not result in a wind, within the framework of the current steady-state hybrid-MHD-dust-driven theory. In addition, the spallation of alumina dust at a distance of around $\approx 7.53R_0$ results in the discontinuity seen in the lower panel of Figure~\ref{fig:figure4}. Such large discontinuities are a source of concern, particularly because they can lead to fluid flow instabilities and shocks. However, this is not within the realm of investigation of the current theory, a simple steady-state treatment. It may be possible that such instabilities may precipitate in motion of material back towards the photosphere as is seen in \cite[][]{Lobel2001}. Thus we see that scenarios 3a and 3b are theoretically possible; however, the primary concern is that they predict sub-Alfv\'enic velocities at large distances. Thus the results of Scenarios 3a and 3b cannot be completely reconciled with observations. This may well point towards the inference that the simplest scenario discussed earlier, Scenario 1a, may well the be the most pertinent.

We now turn our attention to the final scenario before summarising the results obtained in this study.

\subsection{Scenario 4: Alumina spallation close to the photosphere}

Presently, we discuss the cases where alumina dust condensation occurs in both small and large quantities close to the photosphere and subsequent spallation occurs at around $1.5R_0$; Scenarios 4a and 4b respectively. The rationale being, that the reason why alumina dust is not seen between $1.5R_0$ and around $30R_0$ is that it gets destroyed at around $1.5R_0$. This is an alternative to assuming that the alumina grains are transparent until they accumulate silicates on their surfaces. The question then is whether it is possible to achieve a hybrid outflow with alumina dust formation and spallation included in the framework of the hybrid MHD-dust-driven picture. It is to be mentioned at the very outset that the formation of alumina in both small and large quantities was not seen to produce any appreciable influence on the stellar efflux in this scenario. The silicate dust and corresponding gas velocity profiles obtained were seen to be nearly identical to those shown in Figure~\ref{fig:figure1}. Thus, formation and subsequent spallation of alumina, all within $r=1.5R_0$ was not seen to be a physically relevant picture for the stellar efflux of Betelgeuse. Nevertheless, a brief description of the calculation and some pertinent points are conveyed below.

For Scenarios 4a and 4b it was assumed that alumina condensation occurs at $r=1.38R_0$ in small and large quantities. Thus Scenario 4a has parameters $\Gamma_d^\mathrm{alumina}=0.05$ and $\langle \delta_\mathrm{alumina} \rangle =1/20000$, while Scenario 4b has parameters $\Gamma_d^\mathrm{alumina}=0.5$ and $\langle \delta_\mathrm{alumina} \rangle =1/2000$. From $r=R_0$ to $r=1.38R_0$, Eq.~(\ref{eq:1}) is integrated as a pure WD wind without the dust parameter. It is to be mentioned that these parts of the solutions are not part of the WD critical solution. In the region $1.38R_0 \leq r \leq 1.5R_0$, the winds are hybrid MHD-dust-driven winds and Eq.~(\ref{eq:1}) is integrated with the presence of dust parameters for the two models. However alumina is dust is then abruptly destroyed at a distance of $1.5R_0$ and therefore in the region $1.5R_0 < r < 30R_0$ the wind is a pure WD type of wind and Eq.~(\ref{eq:1}) is integrated without the dust parameter. The portion of the solution in the region $1.5R_0 < r < 30R_0$ corresponds to a pure WD critical solution, one that passes through all three critical points. The critical solution then carries the wind out to a distance of $30R_0$ at which point, silicate condensation occurs and silicate dust grains form. The dust and gas velocities for the two models in Scenario 4a and 4b in the region $r \leq 1.5R_0$ were calculated to be miniscule; $\leq 10^{-10}$~cm/s and as a result immeasurable and not detectable in the form of wind signatures. Thus in this scenario, for all intents and purposes, alumina dust condensation can be considered to be absent in the wind as it does not produce any appreciable effect, in which case Scenario 4 quite simply reduces to Scenario 1; the most viable picture thus far in the discussion.

This brings us to the end of our discussion of the results obtained in this study. In the following section, the findings are summarised.

\section{Conclusion}
In this paper we presented a hybrid-MHD-dust-driven wind model for the red supergiant
Betelgeuse ($\alpha-$Orionis). The model is a direct application of
our previously derived theory that consists of incorporating a
dust-driven wind with a Weber-Davis MHD equatorial wind
\citep[see][]{Thirumalai2010}.

Overall, the results shown above indicate that MHD effects
and radiation pressure on dust grains can have a
complementary role to play in the winds of supergiants such as
Betelgeuse, alongside the altogether complicated and involved physics
of thermal pulsation and convection and other equally complex phenomena such as MHD or acoustic waves.

Within the framework of the model, we investigated four different scenarios for dust formation in the atmosphere of Betelgeuse. It was seen that the simplest  hybrid wind scenario; Scenario 1a, was perhaps the most viable one as well. In this picture, a pure WD wind was assumed to begin
at the surface of the star, one that would eventually leave the star
as a hybrid-MHD-dust-driven wind after the formation of dust grains at
the dust condensation radius at around $30R_0$. This provided a mechanism for lifting
stellar material from the photosphere of the star up into the
atmosphere while maintaining low Mach numbers for the
wind in the inner wind region. The dust formation was assumed to occur abruptly at the
dust condensation radius where the temperature is low enough for silicate grains to condense. In our
model, all the dust grains were assumed to be perfectly spherical with
identical size. It was implicitly assumed that radiation pressure was
purely in the radial direction without scattering. The opacity of the
grains were implicitly assumed to be such that all of the radiation
impinging on the grains was absorbed and imparted momentum to the
grains. The resulting drag force was assumed to be purely radial as
well.

In Scenario 1a, since dust grain condensation occurs outside the fast point,   
as in our earlier work \citep[see][]{Thirumalai2010}, it was seen that
adopting a range of value for values the dust parameter $\Gamma_d$, resulted in
different hybrid winds with different terminal velocities. 

It was also seen that by adopting the values
for different parameters for Betelgeuse, as given in Table~1, the
resulting hybrid wind model is able to rationalise several of the
observed features of this red supergiant, such as the predicted wind
velocities, observed atmospheric temperatures at different distances
from the photosphere, observed region of dust formation and
temperature inhomogeneities in the circumstellar envelope.

We additionally investigated the possibility of having more than one dust species form in the wind. Scenarios 2-4 dealt with the formation of alumina dust close to the photosphere of Betelgeuse. We also investigated the effect on the efflux, should alumina be destroyed at some distance. In each of these scenarios it was seen that it was possible to form a hybrid-MHD-dust-driven wind, however it was not possible to reconcile the predictions of these models completely with the current observations regarding dust formation and lack of wind signatures interior to the dust shell of Betelgeuse. In addition, we found that for these scenarios to work, within the framework of the hybrid-MHD-dust-driven wind theory, it was required that the surface temperature should be around $T_0=2600~$K, to overlap with recent observations for the close circumstellar environment of Betelgeuse \citep[e.g.][]{Perrin2007}. Therefore, it is to be concluded that while it may well be possible for alumina to form close to the photosphere of Betelgeuse, it probably does not influence the wind outflow interior to $\sim 30R_0$. However, it is to be remembered that the current model is merely a steady-state treatment of an infinitely more complicated picture. All that can be said at this stage is that the current theory seems to point towards the simplest case; Scenario 1a, as the most likely candidate. It is to be mentioned that the current model may play a complementary role alongside the altogether different and complex mechanisms for stellar outflow involving MHD or acoustic waves.  

Future work will extend the modelling towards
a more realistic full two- or three-dimensional picture, concomitantly
including a poloidal magnetic field and allowing the entire system to
be described dynamically with time varying magnetic and velocity
fields with convection. The final and indeed ultimate picture would of
course, be to then include the stochastic effects of dust formation
and dynamics, alongside calculation of the radiative properties of the dust, as is done in some recent
work \citep[see][]{Woitke2005, Lobel2008}. Realisation of this task will
ultimately explore the true nature of the outflow from Betelgeuse and
other red supergiants, which is now beyond doubt, understood to be
more complicated than at first imagined and yet, presenting a hurdle
surmountable in steps. Perhaps the very first of those steps, is the
steady-state model described in the current study.

\section*{Acknowledgements}
This research was supported by funding from NSERC.  The calculations
were performed on computing infrastructure purchased with funds from
the Canadian Foundation for Innovation and the British Columbia
Knowledge Development Fund. The authors are grateful to the referee, Dr. G.M.~Harper for invaluable insight and direction.

\bibliographystyle{mn2e}
\bibliography{betelgeuse}
\label{lastpage}
\end{document}